\documentclass[11pt]{article}

\usepackage{acl}

\usepackage{times}
\usepackage{latexsym}

\usepackage[T1]{fontenc}
\usepackage[utf8]{inputenc}
\usepackage{microtype}
\usepackage{inconsolata}
\usepackage{graphicx}

\usepackage{amsmath}
\usepackage{amssymb}
\usepackage{booktabs}
\usepackage{hhline}
\usepackage{subcaption}
\usepackage{xcolor}
\usepackage{multirow}
\usepackage{colortbl}
\usepackage{arydshln}
\usepackage{enumitem}
\usepackage{tcolorbox}
\usepackage{pgffor}
\usepackage{arydshln}
\usepackage{xspace}
\usepackage{mdframed}

\newcommand{\ourmethod}{LICD\xspace}

\title{Why Thinking Hurts: Diagnosing and Rectifying Linguistic Inertia in Large Language Models for Recommendation}

\author{%
   Luankang Zhang\textsuperscript{1},
   Yonghao Huang\textsuperscript{1},
   Hang Lv\textsuperscript{1},
   Xuyang Zhi\textsuperscript{1},
   Mingjia Yin\textsuperscript{1},\\
   \textbf{Yuyang Ye}\textsuperscript{1},
   \textbf{Wei Guo}\textsuperscript{2},
   \textbf{Hao Wang}\textsuperscript{1}\footnotemark[1],
   \textbf{Enhong Chen}\textsuperscript{1} \\
   \textsuperscript{1}University of Science and Technology of China \quad
   \textsuperscript{2}Huawei Technologies
}

\begin{document}
\maketitle

\begingroup
\renewcommand{\thefootnote}{\fnsymbol{footnote}}
\footnotetext[1]{Corresponding authors. Contact wanghao3@ustc.edu.cn}
\endgroup
\setcounter{footnote}{0}

\begin{abstract}
Chain-of-Thought (CoT) reasoning is widely used to improve LLM performance, and recent foundation recommender models adopt it by generating textual reasoning before predicting target items represented by Semantic IDs (SIDs). However, we observe that enabling thinking mode in models such as OpenOneRec can degrade recommendation quality by up to 25\%. We investigate this failure and identify \textit{Linguistic Inertia}: when a textual CoT segment is inserted before SID generation, the model relies more on natural-language context and less on historical SID evidence. Further analyses show that this effect is amplified by reduced access to historical information and longer CoT lengths. To mitigate it, we propose \textit{Linguistic-Inertia-Calibrated Decoding} (LICD), a training-free framework that combines \textit{Reasoning-Chain Compression} and \textit{Bias-Subtracted Contrastive Inference}. Experiments on three large-scale benchmarks show that LICD consistently outperforms both no-thinking and original-thinking baselines. Our code is available at \url{https://github.com/USTC-StarTeam/LICD}.
\end{abstract}

% \section{Introduction}
\section{Introduction}
\label{sec:introduction}

Large language models (LLMs) have demonstrated strong reasoning abilities through Chain-of-Thought (CoT) prompting, where step-by-step reasoning improves their performance on a wide range of tasks~\citep{wei2022chain, guo2025deepseek}.
A natural expectation is that such reasoning should also benefit recommender systems.
By explaining why a user may prefer certain items, the model may form a more coherent understanding of user intent before making a prediction.
Recent foundation recommender models~\citep{zhou2025openonerec, liu2025onerec} have adopted this paradigm: a unified LLM first generates a textual reasoning chain and then predicts the target item represented by hierarchical Semantic IDs (SIDs).

\begin{figure}[t]
    \centering
    \includegraphics[width=0.9\linewidth]{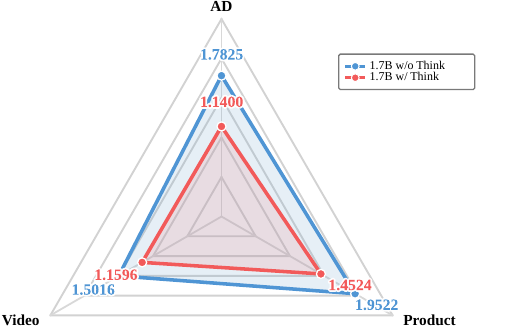}
    \caption{Impact of Thinking Mode on OpenOneRec.}
    \label{fig:think}
\end{figure}

However, we observe a counter-intuitive phenomenon: enabling thinking mode can sometimes reduce recommendation quality.
As shown in Fig.~\ref{fig:think}, on three large-scale benchmarks, activating CoT reasoning in OpenOneRec leads to a performance drop of up to 25\% compared with direct prediction without reasoning.
This observation raises a question for both NLP and recommender systems:
\textit{Why can reasoning be harmful, and can we make it beneficial for SID prediction?}

We investigate this question through an empirical analysis and identify a phenomenon we call \textbf{Linguistic Inertia}.
\textit{It refers to the tendency of an LLM to rely more on natural-language context and less on historical SID evidence} when a textual CoT segment is inserted before SID generation.
In foundation recommender models, this means that the model may underuse the rich information in historical interactions and instead depend more on the reasoning chain.
Since the CoT is not always correct or sufficiently informative, amplified Linguistic Inertia can negatively affect the final recommendation.
Our analysis shows that two factors strengthen this effect: reduced historical information and increased CoT length.

Motivated by these findings, we propose \textit{Linguistic-Inertia-Calibrated Decoding} (LICD), a training-free inference framework that controls the two factors above to prevent textual information from overwhelming the SID history.
LICD consists of two components.
First, \textit{Reasoning-Chain Compression} distills the verbose CoT into a single structured preference statement, reducing textual length while retaining intent-related signals.
The compressed reasoning chain is then used to construct the \textit{Expert} context.
Second, \textit{Bias-Subtracted Contrastive Inference} weakens historical information to construct an \textit{Amateur} context, which exposes the text-dominated behavior caused by Linguistic Inertia.
LICD then calibrates the final ranking by subtracting the normalized Amateur score from the Expert score.
Experiments on three public datasets show that LICD consistently outperforms both the no-thinking and original-thinking baselines.
Our main contributions are as follows:
\begin{itemize}[leftmargin=*]
    \item We identify \textit{Linguistic Inertia}, a systematic phenomenon in thinking-enabled foundation recommender models, where textual reasoning makes SID prediction rely more on natural-language context and less on historical SID evidence.
    
    \item We identify two factors that amplify Linguistic Inertia: reduced historical information and increased CoT length.
    
    \item We propose LICD, an analysis-driven training-free framework that calibrates this reasoning shift through \textit{Reasoning-Chain Compression} and \textit{Bias-Subtracted Contrastive Inference}.
    
    \item Extensive experiments show that LICD consistently improves over both thinking and no-thinking baselines, mitigating the negative effect of Linguistic Inertia.
\end{itemize}

\section{Empirical Analysis}
\label{sec:empirical}

In this section, we investigate how the thinking mode changes Semantic-ID prediction in foundation recommender models.
Our analysis reveals a systematic phenomenon we term \textit{Linguistic Inertia}: when a textual CoT segment is inserted before SID generation, the model tends to rely more on the textual reasoning region and less on the historical SID sequence when making the final prediction.
We further analyze the factors that amplify this effect and find that Linguistic Inertia becomes significantly stronger when the available historical information is reduced or when the textual CoT segment becomes longer.

\subsection{Identifying Linguistic Inertia}
\label{sec:linguistic_inertia}

We begin by examining how the textual CoT segment alters the model's decision-making at the critical prediction position---the token where the model outputs the target Semantic ID.

\paragraph{Setup.}
Let $x$ denote the input sequence containing both Semantic ID (SID) tokens from the user's interaction history and textual instruction tokens.
In thinking mode, the model generates a textual CoT $c$ before predicting the target item $y$.
We extract the attention distribution at the prediction position from the last Transformer layer, averaged over all heads, and compare Think=Off, where $y$ is predicted from $x$ alone, with Think=On, where $y$ is predicted from the concatenated context $(x,c)$.

\begin{figure}[t]
    \centering
    \begin{subfigure}[b]{0.48\linewidth}
        \centering
        \includegraphics[width=\linewidth]{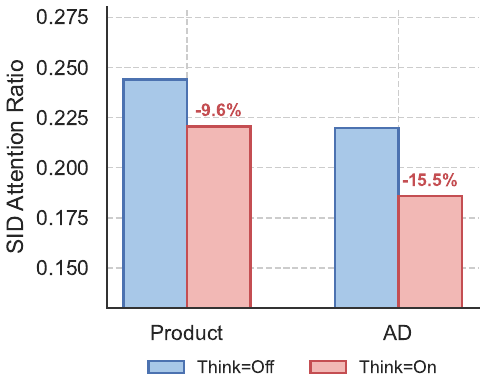}
        \caption{SAR comparison.}
        \label{fig:li_bar}
    \end{subfigure}
    \hfill
    \begin{subfigure}[b]{0.48\linewidth}
        \centering
        \includegraphics[width=\linewidth]{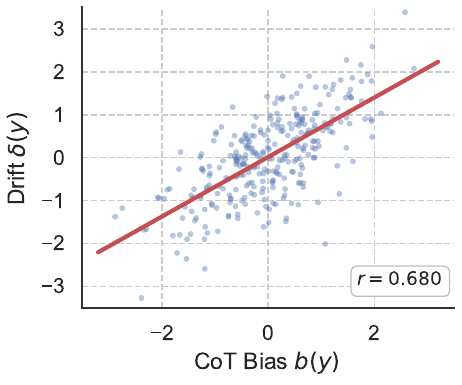}
        \caption{Drift--bias correlation.}
        \label{fig:li_scatter}
    \end{subfigure}
    \caption{Evidence of Linguistic Inertia. (a) The SID Attention Ratio drops when thinking mode is enabled, indicating that attention shifts away from historical SID tokens. (b) The logit change caused by adding CoT is highly correlated with the logit pattern from the CoT-only context ($r=0.680$), suggesting that the final decision is more influenced by textual information.}
    \label{fig:linguistic_inertia}
\end{figure}

\paragraph{Observation 1: Attention shifts from SID history to textual context.}
We define the \textbf{SID Attention Ratio} as the fraction of total attention mass allocated to SID tokens at the prediction position.
Formally, let $\mathbf{a} \in \mathbb{R}^L$ be the attention vector and $\mathcal{I}$ the set of SID token positions:
\begin{equation}
\text{SID Attention Ratio (SAR)} =
\sum_{i \in \mathcal{I}} a_i \bigg/ \sum_{j=1}^{L} a_j.
\label{eq:sar}
\end{equation}
As shown in Fig.~\ref{fig:linguistic_inertia}(a), enabling thinking mode reduces the SID Attention Ratio on both datasets (Product: $-9.6\%$; AD: $-15.5\%$).
This suggests that inserting a textual CoT segment reallocates attention away from historical SID tokens, leaving relatively more attention mass on non-SID textual context at the prediction position.

\paragraph{Observation 2: History+CoT logits become closer to text-only logits.}
The attention analysis shows reduced reliance on SID history; we next examine whether this change is reflected in the SID logit space.
Let $\boldsymbol{\delta}=\ell(y\mid x,c)-\ell(y\mid x)$ denote the logit change caused by inserting the textual CoT into the history context, and let $\mathbf{b}=\ell(y\mid c)-\ell(y\mid x)$ denote the contrast between a CoT-only textual context and the history-only context.
If $\boldsymbol{\delta}$ and $\mathbf{b}$ are highly correlated, then the model's decision under history+CoT is closer to a text-only decision than to a primarily history-based one.

We define the drift--bias correlation as:
\begin{equation}
r = \mathrm{PearsonCorr}(\boldsymbol{\delta}, \mathbf{b}).
\label{eq:drift_bias_corr}
\end{equation}

A larger $r$ means that the effect of adding CoT is more consistent with the CoT-only contrast.
As shown in Fig.~\ref{fig:linguistic_inertia}(b), Product obtains a strong positive correlation ($r=0.680$), indicating that the SID logits under history+CoT are substantially dominated by the textual CoT.

These two observations jointly reveal:

\definecolor{myslategray}{RGB}{47, 79, 79}
\definecolor{lightbluebg}{RGB}{240, 248, 255}

\begin{mdframed}[
  linecolor=myslategray,
  linewidth=1.5pt,
  roundcorner=8pt,
  backgroundcolor=lightbluebg,
  frametitle={Linguistic Inertia},
  frametitlebackgroundcolor=myslategray!80,
  frametitlefontcolor=white,
]
\textbf{Once a textual CoT is fed into the LLM before SID generation, the LLM tends to rely more on information expressed in natural language and less on information encoded by historical SID tokens.}
\end{mdframed}

Although the textual CoT does not always contain wrong information, it may include hallucinated details, over-generalized descriptions, or coarse preference summaries.
The risk is that, when producing the target Semantic ID, the LLM may over-rely on the textual CoT and underuse the SID sequence, allowing text-derived cues to dominate the recommendation even when the history contains more precise item-level evidence.

\begin{figure*}[t]
    \centering
    \begin{subfigure}[b]{0.3\textwidth}
        \centering
        \includegraphics[width=\linewidth]{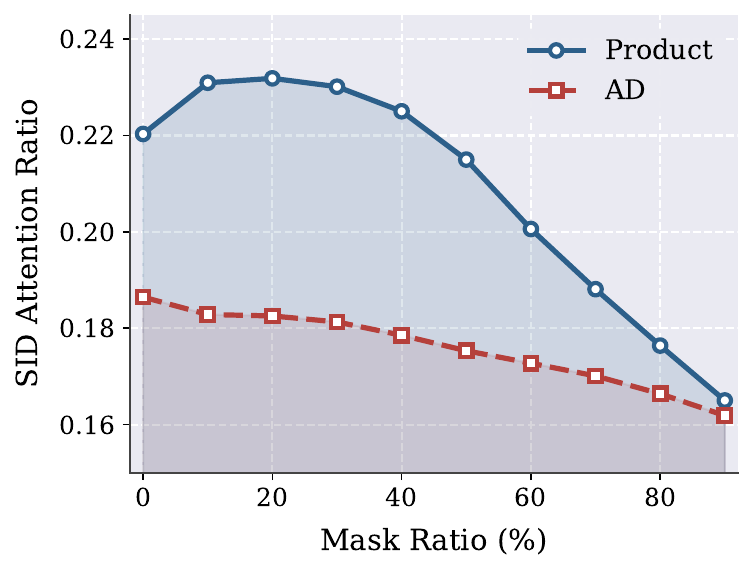}
        \caption{Mask History: SAR}
        \label{fig:mask_hist_sar}
    \end{subfigure}
    \hfill
    \begin{subfigure}[b]{0.3\textwidth}
        \centering
        \includegraphics[width=\linewidth]{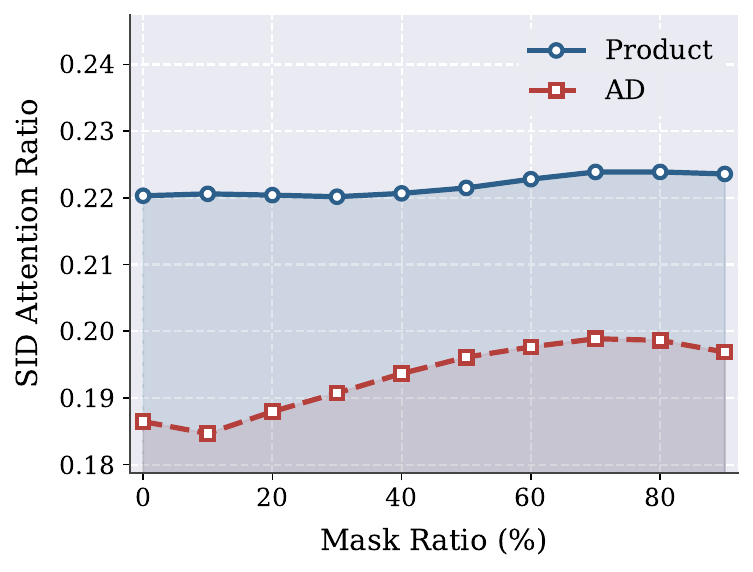}
        \caption{Mask CoT: SAR}
        \label{fig:mask_cot_sar}
    \end{subfigure}
    \hfill
    \begin{subfigure}[b]{0.3\textwidth}
        \centering
        \includegraphics[width=\linewidth]{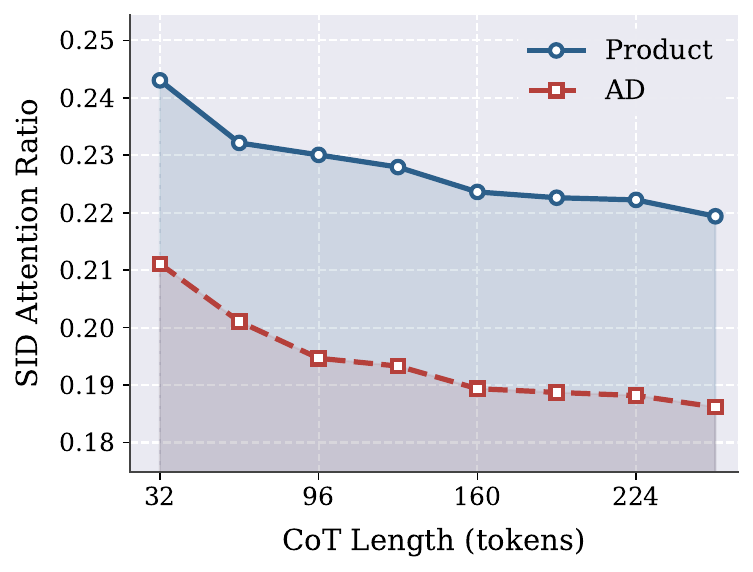}
        \caption{Truncate CoT: SAR}
        \label{fig:cot_length}
    \end{subfigure}
    
    \vspace{0.2em} 
    
    \begin{subfigure}[b]{0.3\textwidth}
        \centering
        \includegraphics[width=\linewidth]{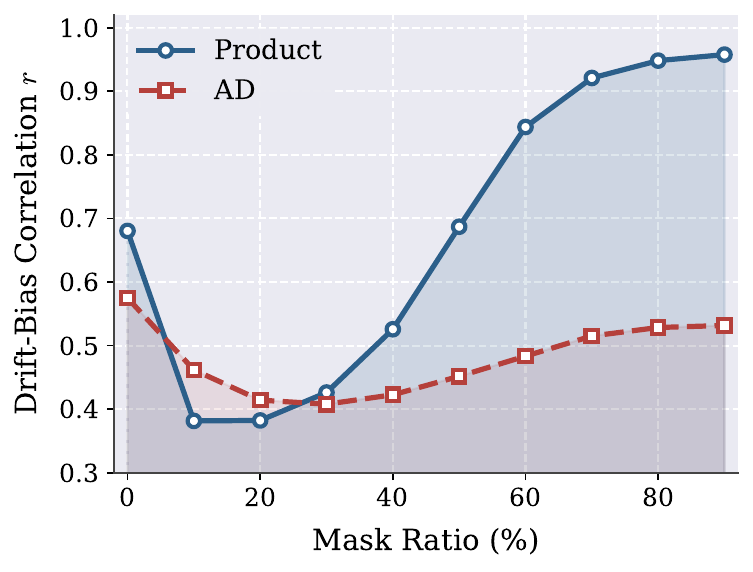}
        \caption{Mask History: $r$}
        \label{fig:mask_hist_r}
    \end{subfigure}
    \hfill
    \begin{subfigure}[b]{0.3\textwidth}
        \centering
        \includegraphics[width=\linewidth]{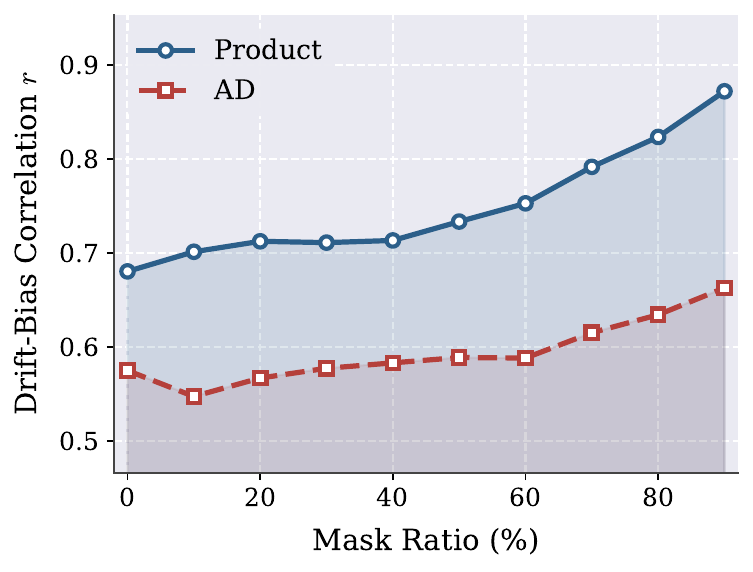}
        \caption{Mask CoT: $r$}
        \label{fig:mask_cot_r}
    \end{subfigure}
    \hfill
    \begin{subfigure}[b]{0.3\textwidth}
        \centering
        \includegraphics[width=\linewidth]{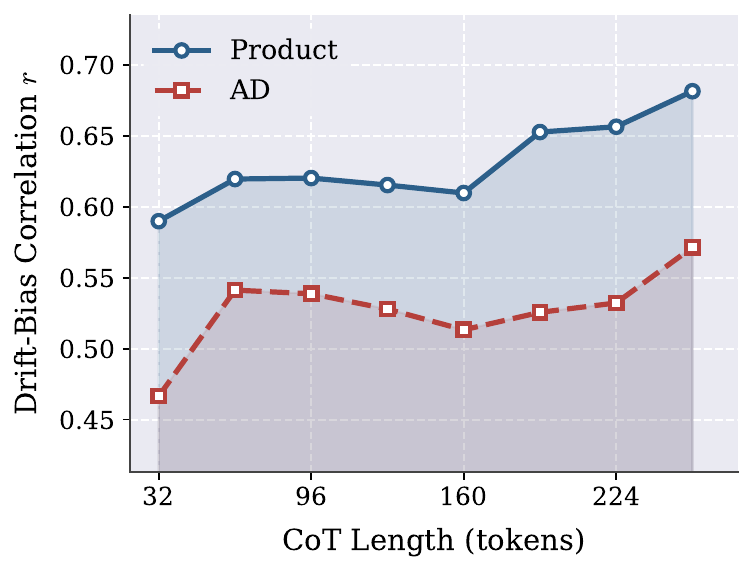}
        \caption{Truncate CoT: $r$}
        \label{fig:cot_length_r}
    \end{subfigure}
    
    \caption{Probing Linguistic Inertia via information masking and length control. \textbf{(a,d)} Masking history SID tokens reduces SID attention and increases $r$, indicating that weakened history makes the decision more text-dominated. \textbf{(b,e)} Masking CoT tokens only mildly restores SID attention while $r$ continues to increase, suggesting that Linguistic Inertia is not solely determined by the semantic content of CoT tokens. \textbf{(c,f)} Truncating CoT shows that textual length is a major contributor to attention dilution. Full details are provided in Appendix~\ref{sec:empirical_appendix}.}
    \label{fig:masking}
\end{figure*}

\subsection{Probing Linguistic Inertia via Information Masking}
\label{sec:masking}

We next examine which factors amplify Linguistic Inertia.
At the prediction position, the LLM conditions on both the historical SID sequence $x$ and the textual CoT segment $c$.
We therefore selectively mask one source at a time: masking history tests whether reduced SID information makes the LLM rely more on the textual CoT, while masking CoT tests whether this effect depends on the readable content of the CoT itself.

\paragraph{Approach.}
For each sample, we generate the textual CoT from the full history and then mask either the historical SID tokens or the CoT tokens with \texttt{<|endoftext|>} at ratios from $0\%$ to $90\%$.
\textbf{Mask History} weakens the SID-history signal while keeping CoT unchanged, whereas \textbf{Mask CoT} weakens the textual CoT while keeping history unchanged.
We keep the sequence length fixed and report SID Attention Ratio and drift--bias correlation $r$.

\paragraph{Results.}
Fig.~\ref{fig:masking} presents three findings.

\textbf{Finding 1: Weakening history amplifies textual dominance.}
When history tokens are progressively masked (Fig.~\ref{fig:masking}(a,d)), the SID Attention Ratio drops and $r$ rises sharply (Product: $0.68 \to 0.96$).
This indicates that as the historical SID sequence becomes less informative, the LLM relies less on history and makes decisions increasingly consistent with the CoT-only context.

\textbf{Finding 2: Masking CoT content does not proportionally restore history reliance.}
When CoT tokens are masked (Fig.~\ref{fig:masking}(b,e)), the SID Attention Ratio increases only marginally (Product: $+1.5\%$ at 90\% mask), while $r$ increases monotonically (Product: $0.68 \to 0.87$).
Since larger $r$ means stronger consistency with the CoT-only logits, this suggests that the LLM remains strongly influenced by the textual CoT region even when much of its readable content is masked.
Thus, Linguistic Inertia is not solely determined by the specific semantic content of CoT tokens.

\textbf{Finding 3: CoT length is a major contributor to attention dilution.}
We truncate the CoT to different lengths while keeping the history intact.
As shown in Fig.~\ref{fig:masking}(c), the SID Attention Ratio decreases as CoT length increases, and at 32 tokens it nearly recovers to the Think=Off level on Product ($0.243$ vs.\ $0.243$).
Since truncation changes both length and content, this does not rule out semantic effects; however, it shows that the amount of textual CoT is an important factor behind attention dilution.
Fig.~\ref{fig:masking}(f) suggests that shortening CoT restores SID attention more effectively than it removes logit-level text consistency.

\paragraph{Implications for method design.}
The analysis identifies two factors that amplify Linguistic Inertia: reduced history information and increased CoT length.
Finding 3 suggests that the textual reasoning channel should be length-controlled before SID prediction, since long free-form CoT reduces attention to SID tokens.
Finding 1 shows that when the historical SID sequence is weakened, the LLM relies more on the textual CoT, revealing how it behaves under text-dominant conditions.

Therefore, our method controls these two factors separately.
We use a short, structured Expert context to limit the textual CoT length, and a weakened-history Amateur context to expose the text-dominant behavior for contrastive calibration.
The goal is not to discard CoT, but to prevent textual information from overwhelming the SID history during final ranking.

\section{Method: Linguistic-Inertia-Calibrated Decoding}
\label{sec:method}

Motivated by the empirical findings above, we propose \textit{Linguistic-Inertia-Calibrated Decoding} (LICD), a training-free inference framework for SID prediction.
The key idea is to preserve the useful preference information introduced by thinking while preventing the textual reasoning segment from overwhelming the historical SID evidence.
To this end, LICD calibrates the final ranking through two complementary contexts: an \textit{Expert} context that uses the full history with a compact preference control, and an \textit{Amateur} context that weakens the history to expose the text-dominated behavior caused by Linguistic Inertia.

LICD consists of two components.
First, \textit{Reasoning-Chain Compression with Entropy Gating} converts the free-form CoT into a short, structured preference statement, and rejects uncertain compressed controls.
This reduces attention dilution while retaining preference-relevant cues.
Second, \textit{Bias-Subtracted Contrastive Inference} constructs a weakened-history Amateur context to expose Linguistic Inertia, and subtracts its normalized score from the Expert score during inference.

\subsection{Reasoning-Chain Compression}
\label{sec:cot-compression}

Section~\ref{sec:masking} shows that longer textual CoT reduces the SID Attention Ratio even when the historical SID sequence is intact. 
This indicates that verbose reasoning can dilute SID-grounded evidence during prediction. 
However, the CoT may still contain useful high-level preference cues. 
We therefore compress the free-form reasoning chain into a short and structured preference control before final SID prediction.

\paragraph{Compression operator.}
Given the interaction history $x$, the model first generates a free-form reasoning chain $c$ under thinking mode. 
We then apply a compression operator:
\begin{equation}
    \hat{c} = \mathcal{T}(c),
    \qquad
    \hat{c} \in \mathcal{C}_{\mathrm{pref}},
    \label{eq:compression_operator}
\end{equation}
where $\mathcal{C}_{\mathrm{pref}}$ is a restricted space of concise preference statements. 
The compressed control $\hat{c}$ is constrained by a fixed template and a length budget:
\begin{equation}
    |\hat{c}| \leq L_c.
    \label{eq:compression_budget}
\end{equation}
Instead of preserving the reasoning trajectory, $\hat{c}$ keeps only the preference-relevant information needed for SID prediction, reducing the verbose linguistic artifacts that amplify Linguistic Inertia.

\paragraph{Instantiation.}
We instantiate $\mathcal{T}$ with an off-the-shelf instruction-following model. 
The compressor removes intermediate reasoning and keeps preference-relevant information from the CoT. 
The full prompt is shown in Fig.~\ref{fig:prompt_structure}. 
It outputs one sentence in the format: 
\textit{The current user's preference is [summary content].}
The sentence is used as the compact control $\hat{c}$ in the Expert context.

\subsection{Bias-Subtracted Contrastive Inference}
\label{sec:contrastive}

Reasoning-chain compression reduces the length of the textual context, but it does not explicitly remove the text-dominated bias identified in Section~\ref{sec:masking}. 
The masking analysis shows that weakening historical SID tokens makes the prediction increasingly aligned with the CoT-only decision pattern. 
We therefore construct an \textit{Amateur} context with degraded history to expose Linguistic Inertia, and subtract its signal from an \textit{Expert} context that uses the full history and compressed reasoning.

\paragraph{Expert context.}
The Expert context conditions on the full interaction history $x$ and the compressed preference control $\hat{c}$:
\begin{equation}
    z_E(y) = \log P_{\theta}(y \mid x, \hat{c}),
    \qquad y \in \mathcal{Y}.
    \label{eq:zE}
\end{equation}
This context preserves SID-grounded evidence while retaining the compact preference cues extracted from the reasoning chain.

\paragraph{Amateur context.}
To capture the text-dominated bias, we weaken the history by masking a fraction $\rho$ of SID tokens:
\begin{equation}
    x_{\rho} = \mathrm{Mask}_{\rho}(x).
    \label{eq:masked_history}
\end{equation}
We then regenerate a free-form reasoning chain from the degraded history,
\begin{equation}
    c_A \sim P_{\theta}(\cdot \mid x_{\rho}),
    \label{eq:amateur_cot}
\end{equation}
and compute the Amateur score:
\begin{equation}
    z_A(y) = \log P_{\theta}(y \mid x_{\rho}, c_A).
    \label{eq:zA}
\end{equation}
Since the SID evidence in $x_{\rho}$ is weakened, high-probability candidates under this context are more likely to reflect Linguistic Inertia rather than precise history-grounded preference.

\paragraph{Score normalization.}
Because the two contexts may have different score scales, we normalize scores within the candidate set $\mathcal{Y}$:
\begin{equation}
    \tilde{z}_k(y)
    =
    \frac{z_k(y)-\mu_k}{\sigma_k+\epsilon},
    \qquad k \in \{E,A\},
    \label{eq:zscore}
\end{equation}
where $\mu_k$ and $\sigma_k$ are the mean and standard deviation of $\{z_k(y)\}_{y\in\mathcal{Y}}$, and $\epsilon$ is a small constant.

\paragraph{Bias-subtracted scoring.}
The ranking score is:
\begin{equation}
    S(y)
    =
    \tilde{z}_E(y)
    -
    \alpha \cdot \tilde{z}_A(y),
    \label{eq:final_score}
\end{equation}
where $\alpha \geq 0$ controls the strength of bias subtraction. 
The final prediction is:
\begin{equation}
    \hat{y}
    =
    \arg\max_{y\in\mathcal{Y}} S(y).
    \label{eq:final_prediction}
\end{equation}

This contrastive score rewards candidates supported by the full-history Expert context and penalizes candidates favored by the weakened-history Amateur context. 
In this way, the model can use reasoning-derived preference cues while reducing the Linguistic Inertia on the final SID ranking.

\section{Experiments}
\subsection{Experimental Setup}
\label{sec:exp_setup}

\begin{table*}[t]
\centering
\renewcommand{\arraystretch}{1.25}
\setlength{\tabcolsep}{3.5pt}
\resizebox{\textwidth}{!}{%
\begin{tabular}{@{} cl cccc cccc cccc @{}}
\toprule
& & \multicolumn{4}{c}{\textbf{AD}}
& \multicolumn{4}{c}{\textbf{Product}}
& \multicolumn{4}{c}{\textbf{Video}} \\
\cmidrule(lr){3-6}
\cmidrule(lr){7-10}
\cmidrule(lr){11-14}
& \textbf{Method}
& R@5 & R@10 & N@5 & N@10
& R@5 & R@10 & N@5 & N@10
& R@5 & R@10 & N@5 & N@10 \\
\midrule

% ===== Discriminative Recommender Models =====
& SASRec
& 0.3609 & 0.6689 & 0.3403 & 0.4570
& 0.1701 & 0.3038 & 0.2365 & 0.2451
& 0.1881 & 0.3649 & 0.2932 & 0.3472 \\
& HSTU
& 0.4593 & 0.8548 & 0.4337 & 0.5848
& 0.1862 & 0.3491 & 0.2352 & 0.2756
& 0.2029 & 0.3886 & 0.3224 & 0.3552 \\
\multirow{-3}{*}{\rotatebox[origin=c]{90}{\scriptsize Disc.}}
& ReaRec
& 0.4121 & 0.7207 & 0.3959 & 0.5121
& 0.2211 & 0.3954 & 0.2814 & 0.3183
& 0.1967 & 0.4068 & 0.3110 & 0.3653 \\
\midrule

% ===== Generative Recommender Models =====
& TIGER
& 0.3973 & 0.6742 & 0.4109 & 0.5107
& 0.1258 & 0.2287 & 0.1478 & 0.1795
& 0.2288 & 0.4133 & 0.3920 & 0.4090 \\
\multirow{-2}{*}{\rotatebox[origin=c]{90}{\scriptsize Gen.}}
& LC-Rec
& 0.5796 & 1.0047 & 0.5730 & 0.7271
& 0.2765 & 0.5185 & 0.3468 & 0.4141
& 0.1568 & 0.3637 & 0.2627 & 0.3202 \\
\midrule

% ===== LLM-as-Rec: 1.7B =====
& \textit{ThinkOff}
& 1.7825 & 2.9611 & 1.7210 & 2.1437
& 1.9522 & 3.0202 & 2.4071 & 2.5966
& 1.1805 & 2.0120 & 1.9114 & 1.9161 \\
& \textit{ThinkOn}
& 1.1400 & 1.9996 & 1.0993 & 1.4090
& 1.4524 & 2.2759 & 1.7974 & 1.9510
& 1.1596 & 1.9994 & 1.9795 & 1.9749 \\
& \textbf{LICD}
& \textbf{1.9104} & \textbf{3.1701}
& \textbf{1.8791} & \textbf{2.3212}
& \textbf{2.1307} & \textbf{3.3102}
& \textbf{2.6949} & \textbf{2.8921}
& \textbf{1.2068} & \textbf{2.0420}
& \textbf{2.0170} & \textbf{2.0089} \\
\rowcolor{gray!15}
\cellcolor{white}
\multirow{-4}{*}{\rotatebox[origin=c]{90}{\scriptsize OpenOneRec-1.7B}}
& $\Delta$%
& \small +7.18 & \small +7.06 & \small +9.19 & \small +8.28
& \small +9.14 & \small +9.60 & \small +11.96 & \small +11.38
& \small +2.23 & \small +1.49 & \small +1.89 & \small +1.72 \\
\midrule

% ===== LLM-as-Rec: 8B =====
& \textit{ThinkOff}
& 2.3267 & 3.9311 & 2.2941 & 2.8560
& 2.4274 & 3.7792 & 3.0245 & 3.2798
& 1.5016 & 2.0420 & 2.9910 & 2.4629 \\
& \textit{ThinkOn}
& 1.2301 & 2.0836 & 1.1957 & 1.4965
& 1.7072 & 2.6095 & 2.1238 & 2.2785
& 1.4963 & 2.0184 & 2.9273 & 2.4194 \\
& \textbf{LICD}
& \textbf{2.4891} & \textbf{4.0424}
& \textbf{2.3764} & \textbf{2.9316}
& \textbf{2.5423} & \textbf{3.9019}
& \textbf{3.1035} & \textbf{3.3727}
& \textbf{1.5417} & \textbf{2.0960}
& \textbf{3.0785} & \textbf{2.5365} \\
\rowcolor{gray!15}
\cellcolor{white}
\multirow{-4}{*}{\rotatebox[origin=c]{90}{\scriptsize OpenOneRec-8B}}
& $\Delta$%
& \small +6.98 & \small +2.83 & \small +3.59 & \small +2.65
& \small +4.73 & \small +3.25 & \small +2.61 & \small +2.83
& \small +2.67 & \small +2.64 & \small +2.93 & \small +2.99 \\
\bottomrule
\end{tabular}%
}
\caption{Main results on three datasets. R@$K$ and N@$K$ denote
Recall@$K$ and NDCG@$K$, respectively. All metric values are multiplied
by 100. $\Delta$ reports the relative improvement (\%) of \textbf{LICD}
over OpenOneRec without thinking. Improvements are statistically
significant at $p < 0.05$.}
\label{tab:main_results}
\end{table*}

\paragraph{Datasets and Metrics.}
We evaluate on three large-scale industrial recommendation datasets from RecIF-Bench~\cite{zhou2025openonerec}\footnote{\url{https://huggingface.co/datasets/OpenOneRec/OpenOneRec-RecIF}}: \textbf{AD}, \textbf{Product}, and \textbf{Video}, covering advertising click prediction, e-commerce purchase prediction, and short-video engagement prediction, respectively.
Items are represented as hierarchical Semantic ID triplets $\langle s_a, s_b, s_c \rangle$, where each level has an 8,192-token vocabulary.
User histories longer than 100 interactions are truncated to the most recent 100 interactions.
We report Recall@$K$ and NDCG@$K$ for $K \in \{5,10\}$ on the official test splits.
Dataset statistics are provided in Appendix~\ref{app:exp_setup}.

\paragraph{Baselines.}
We compare with three categories of recommendation models.
\textit{(i) Discriminative recommenders}: \textbf{SASRec}~\cite{kang2018self}, \textbf{HSTU}~\cite{zhai2024actions}, and \textbf{ReaRec}~\cite{tang2025think}.
\textit{(ii) Generative recommenders}: \textbf{TIGER}~\cite{rajput2023recommender} and \textbf{LC-Rec}~\cite{zheng2024adapting}.
\textit{(iii) Foundation recommender models}: \textbf{OpenOneRec}~\cite{zhou2025openonerec} at both 1.7B and 8B scales.
For OpenOneRec, we evaluate both the non-thinking mode (\textit{ThinkOff}) and the raw reasoning mode (\textit{ThinkOn}).
All non-OpenOneRec baselines are trained or evaluated following their domain-specific benchmark protocols, while OpenOneRec is used as released without additional domain-specific fine-tuning.

\paragraph{Implementation.}
Our method is training-free and is instantiated on OpenOneRec-1.7B and OpenOneRec-8B.
The reasoning-chain compressor is Qwen2.5-7B-Instruct used off-the-shelf, without recommendation-specific training.
All hyperparameters are selected on the validation split and then fixed for test evaluation.
Detailed implementation settings and search ranges are reported in Appendix~\ref{app:exp_setup}.

\subsection{Main Results}
\label{sec:main_results}

Table~\ref{tab:main_results} shows that our method consistently achieves the best overall performance across all datasets and model scales.
First, all discriminative and generative recommenders are substantially outperformed by LLM-as-Rec models such as OpenOneRec, even including LLM-based baselines such as LC-Rec.
This highlights the strong advantage of foundation-scale pre-training for recommendation.

Second, the comparison between OpenOneRec \textit{ThinkOn} and \textit{ThinkOff} shows that directly enabling reasoning often degrades recommendation performance rather than improving it.
This supports our finding in Section~\ref{sec:empirical}: raw CoT can induce linguistic inertia, shifting SID prediction away from interaction-driven evidence toward textual reasoning signals.
In contrast, our method recovers from this degradation and further surpasses the \textit{ThinkOff} baseline, indicating that calibrated reasoning can preserve useful preference cues while suppressing text-dominant bias.
The relatively smaller improvement on Video is likely because the base model often fails to activate thinking mode in this domain, leaving limited reasoning signal and thus less linguistic inertia to correct.

Third, the gains are more pronounced on the 1.7B model than on the 8B model.
This suggests that smaller models are more vulnerable to reasoning-induced attention distraction and therefore benefit more from our correction mechanism.

Finally, since LICD is designed as a training-free inference-time calibration method, its practicality depends on whether the additional compression and contrastive scoring introduce acceptable overhead. We therefore report a latency analysis in Appendix~\ref{sec:efficiency}, showing that the added cost is moderate compared with the \textit{ThinkOn} pipeline.

\subsection{Ablation Study}
\label{sec:ablation}

% \begin{table}[t]
% \centering
% \caption{Ablation study. Values are $\times 100$.}
% \label{tab:ablation}
% \small
% \begin{tabular}{lcccc}
% \toprule
% & \multicolumn{2}{c}{\textbf{AD}} & \multicolumn{2}{c}{\textbf{Product}} \\
% \cmidrule(lr){2-3} \cmidrule(lr){4-5}
% \textbf{Variant} & R@5 & N@5 & R@5 & N@5 \\
% \midrule
% \multicolumn{5}{l}{\textit{OpenOneRec-1.7B}} \\
% --ThinkOn  & 1.140 & 1.099 & 1.452 & 1.797 \\
% --ThinkOff & 1.783 & 1.721 & 1.952 & 2.407 \\
% \midrule
% \multicolumn{5}{l}{\textit{CoT Processing Variants}} \\
% (1) w/o Compression & 1.721 & 1.700 & 2.028 & 2.511 \\
% (2) Short Thinking  & 1.768 & 1.706 & 2.031 & 2.572 \\
% \midrule
% \multicolumn{5}{l}{\textit{Decoding Variants}} \\
% (a) w/o Contrastive & 1.831 & 1.808 & 2.050 & 2.516 \\
% % (b) w/o Z-score Norm & 1.891 & 1.866 & 2.128 & 2.683 \\
% (b) Amateur Only & 1.461 & 1.412 & 1.858 & 2.338 \\
% \midrule
% \textbf{Full (Ours)} & \textbf{1.910} & \textbf{1.879} & \textbf{2.131} & \textbf{2.695} \\
% \bottomrule
% \end{tabular}
% \end{table}

\begin{table}[t]
\centering
\small
\begin{tabular}{lcccc}
\toprule
& \multicolumn{2}{c}{\textbf{AD}} & \multicolumn{2}{c}{\textbf{Product}} \\
\cmidrule(lr){2-3} \cmidrule(lr){4-5}
\textbf{Variant} & R@5 & N@5 & R@5 & N@5 \\
\midrule
\multicolumn{5}{l}{\textit{OpenOneRec-1.7B}} \\
\quad - \textit{ThinkOn}  & 1.140 & 1.099 & 1.452 & 1.797 \\
\quad - \textit{ThinkOff} & 1.783 & 1.721 & 1.952 & 2.407 \\
\midrule  % 用横线将 Baseline 组独立分隔开
\multicolumn{5}{l}{\textit{CoT Processing Variants}} \\
\quad (1) w/o Compression & 1.721 & 1.700 & 2.028 & 2.511 \\
\quad (2) Short Thinking  & 1.768 & 1.706 & 2.031 & 2.572 \\
% \midrule
\addlinespace[0.5ex] 
\hdashline
\addlinespace[0.5ex]
\multicolumn{5}{l}{\textit{Decoding Variants}} \\
\quad (a) w/o Contrastive & 1.831 & 1.808 & 2.050 & 2.516 \\
% \quad w/o Z-score Norm & 1.891 & 1.866 & 2.128 & 2.683 \\
\quad (b) Amateur Only & 1.461 & 1.412 & 1.858 & 2.338 \\
\midrule
\textbf{Full (LICD)} & \textbf{1.910} & \textbf{1.879} & \textbf{2.131} & \textbf{2.695} \\
\bottomrule
\end{tabular}
\caption{Ablation study. Values are $\times 100$.}
\label{tab:ablation}
\end{table}

To evaluate the contribution of each component, we conduct an ablation study on OpenOneRec-1.7B over the AD and Product datasets.
We consider two CoT-processing variants and two decoding variants.
For CoT processing, \textit{(1) w/o Compression} replaces the compressed CoT $\hat{c}$ with raw reasoning chain $c$ in the Expert view, while \textit{(2) Short Thinking} prompts the model to generate a shorter reasoning chain without post-hoc compression.
For decoding, \textit{(a) w/o Contrastive} removes the Amateur penalty by setting $\alpha=0$, while \textit{(b) Amateur Only} ranks candidates using only the Amateur score $z_A$.
We report \textit{ThinkOff} and \textit{ThinkOn} of OpenOneRec-1.7B as reference points.

Table~\ref{tab:ablation} reports the results.
The CoT-processing variants show the necessity of explicit compression.
Variant~(1) underperforms Full, indicating that directly using raw CoT in the Expert view reintroduces linguistic inertia.
Variant~(2) also performs worse than Full, suggesting that simply shortening reasoning is insufficient; the CoT still needs to be distilled into a stable preference control for SID prediction.

The decoding variants verify the role of contrastive calibration.
Variant~(a) improves over raw \textit{ThinkOn} but remains below Full, showing compression alone cannot eliminate the text-dominant bias.
Variant~(b) substantially underperforms Full and \textit{ThinkOff}, confirming the Amateur view exposes linguistic inertia under weakened-history conditions and should serve as a negative reference rather than a standalone predictor.

\subsection{Hyperparameter Sensitivity}
\label{sec:hyperparameter}

\begin{figure}[t]
\centering
\includegraphics[width=0.48\columnwidth]{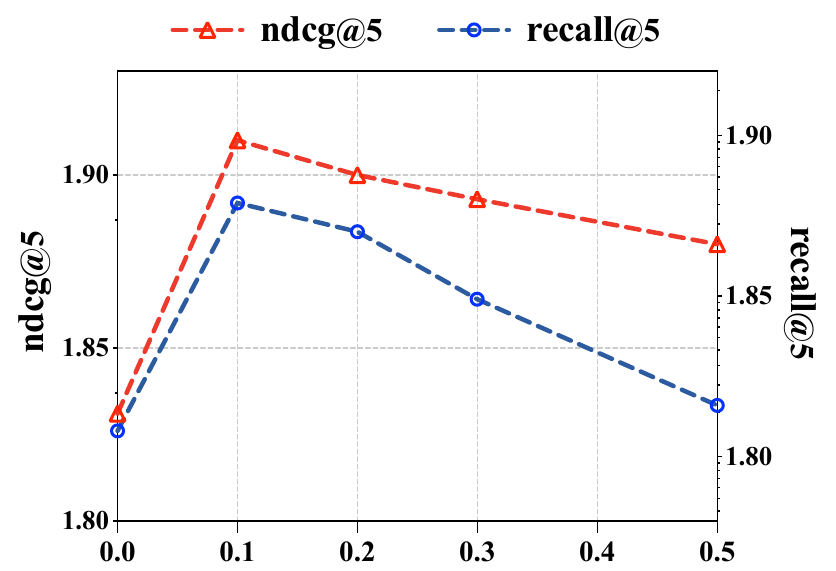}
\hfill
\includegraphics[width=0.49\columnwidth]{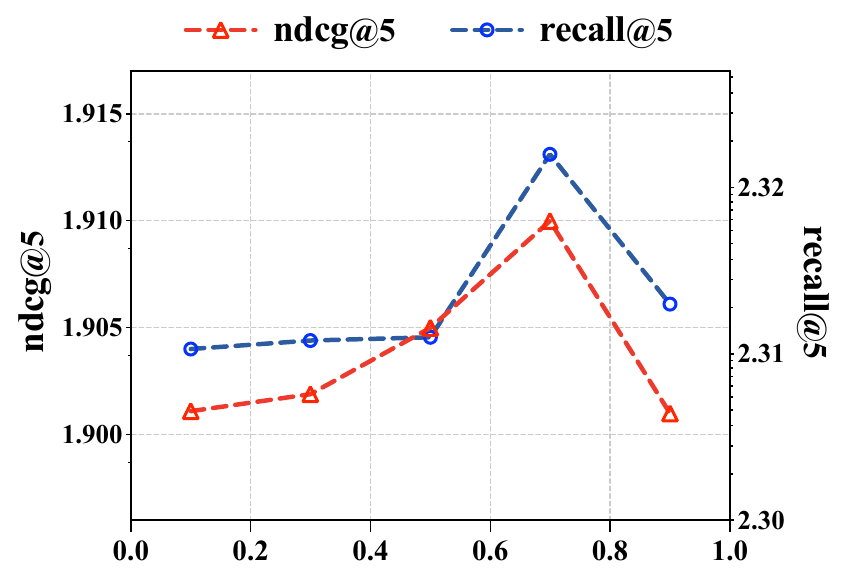}
\caption{Sensitivity analysis on AD with respect to contrastive strength $\alpha$ and Amateur mask ratio $\rho$.}
\label{fig:sensitivity_ad}
\end{figure}

Our method mainly involves two inference-time hyperparameters: the contrastive strength $\alpha$ in Eq.~\ref{eq:final_score} and the mask ratio $\rho$ for constructing the Amateur context in Eq.~\ref{eq:zA}.
Each parameter is swept independently while keeping the others fixed.

\paragraph{Contrastive strength $\alpha$.}

The left panel of Fig.~\ref{fig:sensitivity_ad} shows that a small positive $\alpha$ improves Recall@5 and NDCG@5, indicating the Amateur view provides a useful negative signal for correcting linguistic inertia. When $\alpha$ is too small, the text-dominant bias exposed by the Amateur context is insufficiently removed. As $\alpha$ increases to a moderate range, this bias is better calibrated, and the ranking becomes more reliable. However, large $\alpha$ gradually hurts performance, suggesting that the Amateur distribution may still contain useful coarse preference cues and should not be over-penalized. Overall, the method performs well within a broad small-$\alpha$ range, showing the contrastive strength does not require precise tuning.

\paragraph{Mask ratio $\rho$.}

The right panel of Fig.~\ref{fig:sensitivity_ad} shows that performance remains stable across a wide range of mask ratios.
This suggests that the Amateur context does not rely on a specific masking level.
Instead, it only needs to weaken the historical SID signal enough to reveal the model's text-dominant tendency.
The best performance appears around a moderate-to-large mask ratio, where the history signal is sufficiently weakened while the remaining context can still support meaningful CoT generation.
These results indicate that our method is robust to the choice of $\rho$ and can be applied without extensive per-domain tuning.

\subsection{Layer-wise SID Decomposition}
\label{sec:sid_decomposition}

To further examine how Linguistic Inertia affects hierarchical SID generation, we analyze the impact of ThinkOn at different Semantic ID levels. Each item is represented as a three-level SID triplet $\langle s_a, s_b, s_c \rangle$, where $s_a$, $s_b$, and $s_c$ capture coarse-, middle-, and fine-grained semantics, respectively. Since SIDs are generated autoregressively from coarse to fine, later levels require more precise SID-grounded evidence from user history to distinguish among similar items.

For each level $l \in \{a,b,c\}$, we compare the rank of the ground-truth SID token under \textit{ThinkOff} and \textit{ThinkOn}. An instance is counted as hurt if \textit{ThinkOn} assigns a worse rank than \textit{ThinkOff}, and we report the corresponding hurt rate.

As shown in Table~\ref{tab:sid_layer_hurt}, \textit{ThinkOn} hurts all SID levels, but the degradation becomes stronger at finer levels. On AD, the hurt rate increases from 59.4\% at the coarse level $s_a$ to 68.8\% at the fine level $s_c$. Product shows the same trend, increasing from 58.1\% to 68.0\%. This indicates that raw thinking does not uniformly affect hierarchical SID prediction; instead, its negative effect is amplified as generation moves from coarse semantic prediction to fine-grained item matching.

This trend supports our interpretation that CoT mainly provides coarse-grained preference priors, such as category interests or general consumption intent. Such priors may help narrow the candidate space at the early SID level. However, accurate recommendation further requires selecting the correct fine-grained item within that broad preference direction, which depends more heavily on collaborative signals and item-level evidence encoded in the historical SID sequence. When the textual prior introduced by CoT is over-amplified, the model tends to follow natural-language preference descriptions and underuse the historical SID evidence needed for fine-grained discrimination. Therefore, the ground-truth fine-level token is more likely to be ranked lower under ThinkOn.

These findings further motivate \ourmethod. Rather than discarding CoT, \ourmethod calibrates its influence: reasoning-chain compression preserves useful coarse preference information, while bias-subtracted contrastive inference suppresses the text-dominant shift. In this way, CoT is used as an auxiliary coarse-level prior without overriding SID-grounded evidence for item-level prediction.

\begin{table}[t]
\centering
\small
\begin{tabular}{lccc}
\toprule
\textbf{Dataset} & $s_a$ \textbf{(Coarse)} & $s_b$ \textbf{(Middle)} & $s_c$ \textbf{(Fine)} \\
\midrule
AD      & 59.4 & 64.4 & 68.8 \\
Product & 58.1 & 62.3 & 68.0 \\
\bottomrule
\end{tabular}
\caption{Layer-wise hurt rate of \textit{ThinkOn} compared with \textit{ThinkOff}. Values are percentages.}
\label{tab:sid_layer_hurt}
\vspace{-0.2cm}
\end{table}

\section{Related Work}

\paragraph{\textbf{LLM-Based Recommendation}}
Empowered by deep semantic understanding, Large Language Models (LLMs) have catalyzed a paradigm shift in recommendation systems~\cite{zhou2026survey,pan2025revisiting,ye2025fuxi,xie2025breaking,xu2025multi,wang2025dlf,zhang2025killing,yu2025thought,wang2025universal,zhou2025multi,ye2025fuxibeta,guo2024scaling,zhang2024unified,xie2024breaking,wang2025mf,wang2025generative,zhang2026posi,zhang2026self,ye2026fuxilinear}. Early approaches directly elicited recommendation results via in-context learning or prompt engineering~\cite{achiam2023gpt, gao2023chat, sun2024large, liu2023chatgpt}. To better inject domain-specific collaborative knowledge, instruction tuning emerged as a prevailing paradigm (e.g., P5~\cite{geng2022recommendation}, TALLRec~\cite{bao2023tallrec}), aligning LLMs with recommendation tasks by reformulating user data into natural language sequences. However, verbose text representations struggle to efficiently encode collaborative signals. Consequently, a paradigm shift is occurring towards using discrete Semantic IDs to represent items~\cite{zhou2025openonerec, liu2025onerec, rajput2023recommender, zheng2024adapting}.

\paragraph{\textbf{Reasoning-Enhanced Recommendation}}

In LLM-based recommendation, reasoning capabilities are pivotal for capturing user preferences and generating explainable outcomes~\cite{kim2024pearl, kim2408review, zhou2025hymirec, zhou2025openonerec}. While some explore implicit reasoning~\cite{tang2025think, zhang2025slow}, recent research predominantly prioritizes explicit Chain-of-Thought (CoT) processes. Since raw interaction data lacks reasoning ground truth, the dominant paradigm utilizes capable LLMs as teachers to generate synthetic rationales~\cite{fang2025reason4rec, you2025r, sabouri2025towards, tsai2024leveraging, bismay2025reasoningrec}. To mitigate teacher-induced noise, works like OneRec-think~\cite{liu2025onerec} incorporate retrieval mechanisms for autonomous logical chain construction.

\section{Conclusion}
In this paper, we investigated the failure of CoT reasoning in foundation recommender models. We found that textual reasoning before SID generation introduced \textit{Linguistic Inertia}, which made the model depend more on natural-language context and less on historical SID evidence. We showed that this problem became more serious with less historical information and longer CoT segments. To mitigate it, we proposed LICD, a training-free decoding method that combined reasoning-chain compression with bias-subtracted contrastive inference. Experiments on three large-scale benchmarks demonstrated that LICD outperformed both no-thinking and original-thinking baselines.

\bibliography{Sections/References}

@article{zhang2026posi,
  title={The Next Paradigm Is User-Centric Agent, Not Platform-Centric Service},
  author={Zhang, Luankang and Lv, Hang and Pan, Qiushi and Wang, Kefen and Huang, Yonghao and Miao, Xinrui and Xu, Yin and Guo, Wei and Liu, Yong and Wang, Hao and Chen, Enhong},
  journal={arXiv preprint arXiv:2602.15682},
  year={2026}
}

@article{zhang2026self,
  title={Can Recommender Systems Teach Themselves? A Recursive Self-Improving Framework with Fidelity Control},
  author={Zhang, Luankang and Wang, Hao and Liu, Zhongzhou and Yin, Mingjia and Huang, Yonghao and Li, Jiaqi and Guo, Wei and Liu, Yong and Guo, Huifeng and Lian, Defu and Chen, Enhong},
  journal={arXiv preprint arXiv:2602.15659},
  year={2026}
}

@inproceedings{zhang2024unified,
  title={A unified framework for adaptive representation enhancement and inversed learning in cross-domain recommendation},
  author={Zhang, Luankang and Wang, Hao and Zhang, Suojuan and Yin, Mingjia and Han, Yongqiang and Zhang, Jiaqing and Lian, Defu and Chen, Enhong},
  booktitle={International Conference on Database Systems for Advanced Applications},
  pages={115--130},
  year={2024},
  organization={Springer}
}

@inproceedings{ye2025fuxi,
  title={Fuxi-$\alpha$: Scaling recommendation model with feature interaction enhanced transformer},
  author={Ye, Yufei and Guo, Wei and Chin, Jin Yao and Wang, Hao and Zhu, Hong and Lin, Xi and Ye, Yuyang and Liu, Yong and Tang, Ruiming and Lian, Defu and others},
  booktitle={Companion Proceedings of the ACM on Web Conference 2025},
  pages={557--566},
  year={2025}
}

@article{ye2026fuxilinear,
  title={Fuxi-linear: Unleashing the power of linear attention in long-term time-aware sequential recommendation},
  author={Ye, Yufei and Guo, Wei and Wang, Hao and Zhang, Luankang and Chang, Heng and Zhu, Hong and Ye, Yuyang and Liu, Yong and Lian, Defu and Chen, Enhong},
  journal={arXiv preprint arXiv:2602.23671},
  year={2026}
}

@inproceedings{wang2025universal,
  title={A universal framework for compressing embeddings in ctr prediction},
  author={Wang, Kefan and Wang, Hao and Song, Kenan and Guo, Wei and Cheng, Kai and Li, Zhi and Liu, Yong and Lian, Defu and Chen, Enhong},
  booktitle={International Conference on Database Systems for Advanced Applications},
  pages={84--100},
  year={2025},
  organization={Springer}
}

@article{yu2025thought,
  title={Thought-augmented planning for llm-powered interactive recommender agent},
  author={Yu, Haocheng and Wu, Yaxiong and Wang, Hao and Guo, Wei and Liu, Yong and Li, Yawen and Ye, Yuyang and Du, Junping and Chen, Enhong},
  journal={arXiv preprint arXiv:2506.23485},
  year={2025}
}

@inproceedings{zhang2025killing,
  title={Killing two birds with one stone: Unifying retrieval and ranking with a single generative recommendation model},
  author={Zhang, Luankang and Song, Kenan and Lee, Yi Quan and Guo, Wei and Wang, Hao and Li, Yawen and Guo, Huifeng and Liu, Yong and Lian, Defu and Chen, Enhong},
  booktitle={Proceedings of the 48th International ACM SIGIR Conference on Research and Development in Information Retrieval},
  pages={2224--2234},
  year={2025}
}

@article{wang2025mf,
  title={Mf-gslae: A multi-factor user representation pre-training framework for dual-target cross-domain recommendation},
  author={Wang, Hao and Yin, Mingjia and Zhang, Luankang and Zhao, Sirui and Chen, Enhong},
  journal={ACM Transactions on Information Systems},
  volume={43},
  number={2},
  pages={1--28},
  year={2025},
  publisher={ACM New York, NY}
}

@inproceedings{wang2025generative,
  title={Generative large recommendation models: emerging trends in llms for recommendation},
  author={Wang, Hao and Guo, Wei and Zhang, Luankang and Chin, Jin Yao and Ye, Yufei and Guo, Huifeng and Liu, Yong and Lian, Defu and Tang, Ruiming and Chen, Enhong},
  booktitle={Companion Proceedings of the ACM on Web Conference 2025},
  pages={49--52},
  year={2025}
}

@inproceedings{wang2025dlf,
  title={DLF: Enhancing explicit-implicit interaction via dynamic low-order-aware fusion for CTR prediction},
  author={Wang, Kefan and Wang, Hao and Guo, Wei and Liu, Yong and Lin, Jianghao and Lian, Defu and Chen, Enhong},
  booktitle={Proceedings of the 48th International ACM SIGIR Conference on Research and Development in Information Retrieval},
  pages={2213--2223},
  year={2025}
}

@inproceedings{xu2025multi,
  title={Multi-granularity interest retrieval and refinement network for long-term user behavior modeling in ctr prediction},
  author={Xu, Xiang and Wang, Hao and Guo, Wei and Zhang, Luankang and Yang, Wanshan and Yu, Runlong and Liu, Yong and Lian, Defu and Chen, Enhong},
  booktitle={Proceedings of the 31st ACM SIGKDD Conference on Knowledge Discovery and Data Mining V. 1},
  pages={2745--2755},
  year={2025}
}

@inproceedings{xie2025breaking,
  title={Breaking the Bottleneck: User-Specific Optimization and Real-Time Inference Integration for Sequential Recommendation},
  author={Xie, Wenjia and Wang, Hao and Fang, Minghao and Yu, Ruize and Guo, Wei and Liu, Yong and Lian, Defu and Chen, Enhong},
  booktitle={Proceedings of the 31st ACM SIGKDD Conference on Knowledge Discovery and Data Mining V. 2},
  pages={3333--3343},
  year={2025}
}

@article{ye2025fuxibeta,
  title={Fuxi-$\backslash$beta: Towards a lightweight and fast large-scale generative recommendation model},
  author={Ye, Yufei and Guo, Wei and Wang, Hao and Zhu, Hong and Ye, Yuyang and Liu, Yong and Guo, Huifeng and Tang, Ruiming and Lian, Defu and Chen, Enhong},
  journal={arXiv preprint arXiv:2508.10615},
  year={2025}
}

@article{pan2025revisiting,
  title={Revisiting scalable sequential recommendation with Multi-Embedding Approach and Mixture-of-Experts},
  author={Pan, Qiushi and Wang, Hao and An, Guoyuan and Zhang, Luankang and Guo, Wei and Liu, Yong},
  journal={arXiv preprint arXiv:2510.25285},
  year={2025}
}

@article{zhou2026survey,
  title={A Survey of User Lifelong Behavior Modeling: Perspectives on Efficiency and Effectiveness},
  author={Zhou, Rui and Jia, Qinglin and Chen, Bo and Xu, Peng and Sun, Yijia and Lou, Siyuan and Fu, Chaoxin and Fu, Mengyuan and Shen, Guoming and Zhou, Zheli and others},
  year={2026},
  publisher={Preprints}
}

@inproceedings{zhou2025multi,
  title={MIT: A Multi-Tower Information Transfer Framework Based on Hierarchical Task Relationship Modeling},
  author={Zhou, Rui and Wang, Hao and Guo, Wei and Jia, Qinglin and Xie, Wenjia and Xu, Xiang and Liu, Yong and Lian, Defu and Chen, Enhong},
  booktitle={Companion Proceedings of the ACM on Web Conference 2025},
  pages={651--660},
  year={2025}
}

@article{xie2024breaking,
  title={Breaking determinism: Fuzzy modeling of sequential recommendation using discrete state space diffusion model},
  author={Xie, Wenjia and Wang, Hao and Zhang, Luankang and Zhou, Rui and Lian, Defu and Chen, Enhong},
  journal={Advances in Neural Information Processing Systems},
  volume={37},
  pages={22720--22744},
  year={2024}
}

@article{guo2024scaling,
  title={Scaling new frontiers: Insights into large recommendation models},
  author={Guo, Wei and Wang, Hao and Zhang, Luankang and Chin, Jin Yao and Liu, Zhongzhou and Cheng, Kai and Pan, Qiushi and Lee, Yi Quan and Xue, Wanqi and Shen, Tingjia and others},
  journal={arXiv preprint arXiv:2412.00714},
  year={2024}
}

@article{zhou2025openonerec,
  title={OpenOneRec Technical Report},
  author={Zhou, Guorui and Bao, Honghui and Huang, Jiaming and Deng, Jiaxin and Zhang, Jinghao and She, Junda and Cai, Kuo and Ren, Lejian and Ren, Lu and Luo, Qiang and others},
  journal={arXiv preprint arXiv:2512.24762},
  year={2025}
}

@article{wei2022chain,
  title={Chain-of-thought prompting elicits reasoning in large language models},
  author={Wei, Jason and Wang, Xuezhi and Schuurmans, Dale and Bosma, Maarten and Xia, Fei and Chi, Ed and Le, Quoc V and Zhou, Denny and others},
  journal={Advances in neural information processing systems},
  volume={35},
  pages={24824--24837},
  year={2022}
}

@article{liu2025onerec,
  title={Onerec-think: In-text reasoning for generative recommendation},
  author={Liu, Zhanyu and Wang, Shiyao and Wang, Xingmei and Zhang, Rongzhou and Deng, Jiaxin and Bao, Honghui and Zhang, Jinghao and Li, Wuchao and Zheng, Pengfei and Wu, Xiangyu and others},
  journal={arXiv preprint arXiv:2510.11639},
  year={2025}
}

@article{fang2025reason4rec,
  title={Reason4Rec: Large Language Models for Recommendation with Deliberative User Preference Alignment},
  author={Fang, Yi and Wang, Wenjie and Zhang, Yang and Zhu, Fengbin and Wang, Qifan and Feng, Fuli and He, Xiangnan},
  journal={arXiv preprint arXiv:2502.02061},
  year={2025}
}

@inproceedings{bismay2025reasoningrec,
  title={Reasoningrec: Bridging personalized recommendations and human-interpretable explanations through llm reasoning},
  author={Bismay, Millennium and Dong, Xiangjue and Caverlee, James},
  booktitle={Findings of the Association for Computational Linguistics: NAACL 2025},
  pages={8132--8148},
  year={2025}
}

@inproceedings{bao2023tallrec,
  title={Tallrec: An effective and efficient tuning framework to align large language model with recommendation},
  author={Bao, Keqin and Zhang, Jizhi and Zhang, Yang and Wang, Wenjie and Feng, Fuli and He, Xiangnan},
  booktitle={Proceedings of the 17th ACM conference on recommender systems},
  pages={1007--1014},
  year={2023}
}

@inproceedings{geng2022recommendation,
  title={Recommendation as language processing (rlp): A unified pretrain, personalized prompt \& predict paradigm (p5)},
  author={Geng, Shijie and Liu, Shuchang and Fu, Zuohui and Ge, Yingqiang and Zhang, Yongfeng},
  booktitle={Proceedings of the 16th ACM conference on recommender systems},
  pages={299--315},
  year={2022}
}

@article{achiam2023gpt,
  title={Gpt-4 technical report},
  author={Achiam, Josh and Adler, Steven and Agarwal, Sandhini and Ahmad, Lama and Akkaya, Ilge and Aleman, Florencia Leoni and Almeida, Diogo and Altenschmidt, Janko and Altman, Sam and Anadkat, Shyamal and others},
  journal={arXiv preprint arXiv:2303.08774},
  year={2023}
}

@article{guo2025deepseek,
  title={Deepseek-r1: Incentivizing reasoning capability in llms via reinforcement learning},
  author={Guo, Daya and Yang, Dejian and Zhang, Haowei and Song, Junxiao and Zhang, Ruoyu and Xu, Runxin and Zhu, Qihao and Ma, Shirong and Wang, Peiyi and Bi, Xiao and others},
  journal={arXiv preprint arXiv:2501.12948},
  year={2025}
}

@article{gao2023chat,
  title={Chat-rec: Towards interactive and explainable llms-augmented recommender system},
  author={Gao, Yunfan and Sheng, Tao and Xiang, Youlin and Xiong, Yun and Wang, Haofen and Zhang, Jiawei},
  journal={arXiv preprint arXiv:2303.14524},
  year={2023}
}

@inproceedings{sun2024large,
  title={Large language models for intent-driven session recommendations},
  author={Sun, Zhu and Liu, Hongyang and Qu, Xinghua and Feng, Kaidong and Wang, Yan and Ong, Yew Soon},
  booktitle={Proceedings of the 47th International ACM SIGIR Conference on Research and Development in Information Retrieval},
  pages={324--334},
  year={2024}
}

@inproceedings{zheng2024adapting,
  title={Adapting large language models by integrating collaborative semantics for recommendation},
  author={Zheng, Bowen and Hou, Yupeng and Lu, Hongyu and Chen, Yu and Zhao, Wayne Xin and Chen, Ming and Wen, Ji-Rong},
  booktitle={2024 IEEE 40th International Conference on Data Engineering (ICDE)},
  pages={1435--1448},
  year={2024},
  organization={IEEE}
}

@article{liu2023chatgpt,
  title={Is chatgpt a good recommender? a preliminary study},
  author={Liu, Junling and Liu, Chao and Zhou, Peilin and Lv, Renjie and Zhou, Kang and Zhang, Yan},
  journal={arXiv preprint arXiv:2304.10149},
  year={2023}
}

@article{rajput2023recommender,
  title={Recommender systems with generative retrieval},
  author={Rajput, Shashank and Mehta, Nikhil and Singh, Anima and Hulikal Keshavan, Raghunandan and Vu, Trung and Heldt, Lukasz and Hong, Lichan and Tay, Yi and Tran, Vinh and Samost, Jonah and others},
  journal={Advances in Neural Information Processing Systems},
  volume={36},
  pages={10299--10315},
  year={2023}
}

@inproceedings{tsai2024leveraging,
  title={Leveraging llm reasoning enhances personalized recommender systems},
  author={Tsai, Alicia and Kraft, Adam and Jin, Long and Cai, Chenwei and Hosseini, Anahita and Xu, Taibai and Zhang, Zemin and Hong, Lichan and Chi, Ed H and Yi, Xinyang},
  booktitle={Findings of the Association for Computational Linguistics: ACL 2024},
  pages={13176--13188},
  year={2024}
}

@inproceedings{you2025r,
  title={{R$^2$ec}: Towards Large Recommender Models with Reasoning},
  author={You, Runyang and Li, Yongqi and Lin, Xinyu and Zhang, Xin and Wang, Wenjie and Li, Wenjie and Nie, Liqiang},
  booktitle={The Thirty-ninth Annual Conference on Neural Information Processing Systems},
  year={2025}
}

@article{kim2408review,
  title={Review-driven personalized preference reasoning with large language models for recommendation. CoRR, abs/2408.06276, 2024. doi: 10.48550},
  author={Kim, Jieyong and Kim, Hyunseo and Cho, Hyunjin and Kang, SeongKu and Chang, Buru and Yeo, Jinyoung and Lee, Dongha},
  journal={arXiv preprint arXiv:2408.06276},
  year={2024}
}

@inproceedings{sabouri2025towards,
  title={Towards Explainable Temporal User Profiling with LLMs},
  author={Sabouri, Milad and Mansoury, Masoud and Lin, Kun and Mobasher, Bamshad},
  booktitle={Adjunct Proceedings of the 33rd ACM Conference on User Modeling, Adaptation and Personalization},
  pages={219--227},
  year={2025}
}

@article{zhou2025hymirec,
  title={HyMiRec: A Hybrid Multi-interest Learning Framework for LLM-based Sequential Recommendation},
  author={Zhou, Jingyi and Chen, Cheng and Zuo, Kai and Xu, Manjie and Fu, Zhendong and Chen, Yibo and Tang, Xu and Hu, Yao},
  journal={arXiv preprint arXiv:2510.13738},
  year={2025}
}

@article{tang2025think,
  title={Think before recommend: Unleashing the latent reasoning power for sequential recommendation},
  author={Tang, Jiakai and Dai, Sunhao and Shi, Teng and Xu, Jun and Chen, Xu and Chen, Wen and Wu, Jian and Jiang, Yuning},
  journal={arXiv preprint arXiv:2503.22675},
  year={2025}
}

@article{zhang2025slow,
  title={Slow Thinking for Sequential Recommendation},
  author={Zhang, Junjie and Zhang, Beichen and Sun, Wenqi and Lu, Hongyu and Zhao, Wayne Xin and Chen, Yu and Wen, Ji-Rong},
  journal={arXiv preprint arXiv:2504.09627},
  year={2025}
}

@article{kim2024pearl,
  title={Pearl: A review-driven persona-knowledge grounded conversational recommendation dataset},
  author={Kim, Minjin and Kim, Minju and Kim, Hana and Kwak, Beong-woo and Chun, Soyeon and Kim, Hyunseo and Kang, SeongKu and Yu, Youngjae and Yeo, Jinyoung and Lee, Dongha},
  journal={arXiv preprint arXiv:2403.04460},
  year={2024}
}

@article{pca1,
  title={Principal components analysis (PCA)},
  author={Ma{\'c}kiewicz, Andrzej and Ratajczak, Waldemar},
  journal={Computers \& Geosciences},
  volume={19},
  number={3},
  pages={303--342},
  year={1993},
  publisher={Elsevier}
}

@article{pca2,
  title={Principal component analysis},
  author={Abdi, Herv{\'e} and Williams, Lynne J},
  journal={Wiley interdisciplinary reviews: computational statistics},
  volume={2},
  number={4},
  pages={433--459},
  year={2010},
  publisher={Wiley Online Library}
}

@inproceedings{kang2018self,
  title={Self-attentive sequential recommendation},
  author={Kang, Wang-Cheng and McAuley, Julian},
  booktitle={2018 IEEE international conference on data mining (ICDM)},
  pages={197--206},
  year={2018},
  organization={IEEE}
}

@article{zhai2024actions,
  title={Actions speak louder than words: Trillion-parameter sequential transducers for generative recommendations},
  author={Zhai, Jiaqi and Liao, Lucy and Liu, Xing and Wang, Yueming and Li, Rui and Cao, Xuan and Gao, Leon and Gong, Zhaojie and Gu, Fangda and He, Michael and others},
  journal={arXiv preprint arXiv:2402.17152},
  year={2024}
}

\appendix
% \section{Additional Heatmap Visualizations}
\section{Detailed Experimental Setup}
\label{app:exp_setup}

\subsection{Evaluation Data Statistics}
\label{app:eval_stats}

We evaluate on the official evaluation splits of the RecIF-Bench benchmark~\cite{zhou2025openonerec}.
The benchmark contains three industrial recommendation domains: \textbf{AD}, \textbf{Product}, and \textbf{Video}.
All three domains encode each item as a hierarchical Semantic ID triplet $\langle s_a, s_b, s_c \rangle$, where each level has a vocabulary size of 8,192.
For all methods, user interaction histories exceeding 100 interactions are truncated to the most recent 100 interactions.
Table~\ref{tab:eval_stats} summarizes the evaluation data statistics.
The three domains differ substantially in catalog size and interaction density, ranging from 79K items in AD to over 1.5M items in Video, thus providing a diverse testbed for evaluating recommendation performance across domains.

\begin{table*}[t]
\small
\begin{tabular}{lrrrrr}
\toprule
\textbf{Domain} & \textbf{\#Users} & \textbf{\#Items} & \textbf{\#Inter.} & \textbf{Avg. Hist.} & \textbf{Avg. Tgt.} \\
\midrule
AD      & 27,677  & 79,185      & 1,050K  & 32.5  & 5.4 \\
Product & 27,910  & 365,228     & 1,890K  & 60.9  & 6.8 \\
Video   & 38,781  & 1,577,337   & 4,224K  & 100.0 & 8.9 \\
\bottomrule
\end{tabular}
\centering
\caption{Statistics of evaluation datasets. Histories exceeding 100 interactions are truncated to the most recent 100. Avg. Hist. denotes the average number of historical interactions per user, and Avg. Tgt. denotes the average number of target items per user.}
\label{tab:eval_stats}
\end{table*}

\subsection{Training Data Statistics}
\label{app:train_stats}

Table~\ref{tab:train_stats} reports the training data statistics for the per-dataset baselines, including discriminative recommender models and generative recommender models.
Specifically, SASRec~\cite{kang2018self}, HSTU~\cite{zhai2024actions}, ReaRec~\cite{tang2025think}, TIGER~\cite{rajput2023recommender}, and LC-Rec~\cite{zheng2024adapting} are independently trained on the corresponding domain-specific training split from RecIF-Bench~\cite{zhou2025openonerec}.
The training sets vary significantly in scale: AD contains 4.2M interactions over 161K items, Product contains 16.1M interactions over 1.79M items, and Video contains 75.7M interactions over 11.59M items.

OpenOneRec~\cite{zhou2025openonerec}, categorized as an LLM-as-Rec model in our experiments, follows a different setting.
It is pre-trained on the full cross-domain corpus released with RecIF-Bench, which contains approximately 120M interactions from 200K users across the AD, Product, and Video domains.
We use the released OpenOneRec checkpoints directly and do not perform domain-specific fine-tuning.

\begin{table*}[t]
\centering
\small
\begin{tabular}{lrrrrr}
\toprule
\textbf{Domain} & \textbf{\#Users} & \textbf{\#Items} & \textbf{\#Inter.} & \textbf{Avg. Hist.} & \textbf{Avg. Tgt.} \\
\midrule
AD      & 110,858  & 161,305      & 4.2M   & 32.6  & 5.4 \\
Product & 113,210  & 1,792,738    & 16.1M  & 135.6 & 6.8 \\
Video   & 156,245  & 11,589,098   & 75.7M  & 475.4 & 8.9 \\
\bottomrule
\end{tabular}
\caption{Training data statistics for per-dataset baselines. Avg. Hist. denotes the average number of historical interactions per user, and Avg. Tgt. denotes the average number of target items per user.}
\label{tab:train_stats}
\end{table*}

\subsection{Baseline Details}
\label{app:baseline_details}

We compare against three groups of recommendation models.

\paragraph{Discriminative recommender models.}
\begin{itemize}
    \item \textbf{SASRec}~\cite{kang2018self} is a self-attentive sequential recommendation model that predicts target items from users' historical interaction sequences.
    \item \textbf{HSTU}~\cite{zhai2024actions} is a sequential transduction model designed for large-scale recommendation.
    \item \textbf{ReaRec}~\cite{tang2025think} enhances sequential recommendation with reasoning-augmented user preference modeling.
\end{itemize}

\paragraph{Generative recommender models.}
\begin{itemize}
    \item \textbf{TIGER}~\cite{rajput2023recommender} formulates recommendation as generative retrieval over semantic item IDs.
    \item \textbf{LC-Rec}~\cite{zheng2024adapting} adapts LLMs for recommendation by incorporating collaborative semantics.
\end{itemize}

\paragraph{LLM-as-Rec models.}
\begin{itemize}
    \item \textbf{OpenOneRec}~\cite{zhou2025openonerec} is an open-source foundation recommendation model trained on the RecIF-Bench cross-domain corpus.
    Unlike per-domain baselines, OpenOneRec is pre-trained across multiple recommendation domains and directly generates recommendations through hierarchical Semantic IDs.
    We evaluate OpenOneRec at both 1.7B and 8B scales.
\end{itemize}

For SASRec, HSTU, ReaRec, TIGER, and LC-Rec, we train one model per domain using the corresponding domain-specific training split.
For OpenOneRec, we directly use the released pre-trained checkpoints without additional domain-specific fine-tuning.

\subsection{Implementation Details and Hyperparameter Search}
\label{app:impl_details}

Our method is training-free and is applied to OpenOneRec-1.7B and OpenOneRec-8B.
The reasoning-chain compressor is implemented with Qwen2.5-7B-Instruct and used off-the-shelf without fine-tuning.
All evaluations are performed on GPUs.

Candidate generation uses beam search with beam width $B=32$.
We report Recall@$K$ and NDCG@$K$ for $K \in \{5,10\}$.

For our method, we conduct a grid search over the following hyperparameters:
\[
\resizebox{\linewidth}{!}{$
\begin{aligned}
\alpha &\in \{0.00, 0.05, 0.10, 0.15, 0.20, 0.30, 0.50, 0.70, 1.00\}, \\
% \tau  &\in \{0.00, 0.10, 0.20, 0.30, 0.50, 0.70, 1.00\}, \\
\rho  &\in \{0.10, 0.30, 0.50, 0.70, 0.90\}.
\end{aligned}
$}
\]
Here, $\alpha$ is the contrastive strength in Eq.~\ref{eq:final_score},
 $\rho$ is the mask ratio for constructing the amateur context.

% \subsection{Prompt for Reasoning-Chain Compression}

\begin{figure}[t]
    \centering
    \begin{tcolorbox}[title=\textbf{System Message}, colback=white, colframe=black!75, coltitle=white, sharp corners=south]
        You are a user profiling expert. Please read the analysis process below and directly summarize the user's core preferences.
        
        \textbf{Requirements:}
        \begin{enumerate}
            \item Ignore filler words such as ``I need to analyze'', ``First'', etc.
            \item Provide only one conclusion sentence.
            \item Format: The current user's preference is [summary content].
        \end{enumerate}
        
        \textbf{Input Example:} ... \textbf{Output Example:} ...
    \end{tcolorbox}
    
    \begin{tcolorbox}[title=\textbf{User Message}, colback=gray!10, colframe=black!75, coltitle=white, sharp corners=north]
        \{Original\_CoT\}
    \end{tcolorbox}
    
    \caption{Prompt structure for reasoning-chain compression. The free-form CoT is compressed into a single structured preference statement.}
    \label{fig:prompt_structure}
\end{figure}

\section{Details of the Empirical Analysis Experiments}
\label{sec:empirical_appendix}

This section provides full details of the empirical analysis experiments in Section~\ref{sec:empirical} to ensure reproducibility.

\paragraph{Model and data.}
All experiments use the OpenOneRec-1.7B model (Qwen3ForCausalLM architecture) with 24,576 Semantic ID tokens organized in a three-level hierarchy (\texttt{<s\_a\_*>}, \texttt{<s\_b\_*>}, \texttt{<s\_c\_*>}, 8,192 tokens each). We evaluate on two datasets: \textbf{Product} (e-commerce item recommendation) and \textbf{AD} (advertisement recommendation), using 300 randomly sampled test instances per dataset.

\paragraph{Attention extraction.}
We extract attention weights from the last Transformer layer at the prediction position (the token immediately following \texttt{<|sid\_begin|>}). Attention is averaged across all heads to obtain a single distribution over the input sequence. The SID Attention Ratio is computed as the sum of attention weights on all SID-related token positions (including \texttt{<|sid\_begin|>}, \texttt{<|sid\_end|>}, and all \texttt{<s\_a/b/c\_*>} tokens) divided by the total attention mass.

\paragraph{Drift--bias correlation.}
For each sample, we compute logits over all 8,192 first-level SID tokens (\texttt{<s\_a\_*>}) under three conditions: (1) history only ($x$), (2) history + CoT ($x, c$), and (3) CoT only (history replaced with a generic instruction). The drift vector $\boldsymbol{\delta}$ and bias vector $\mathbf{b}$ are computed in this 8,192-dimensional logit space, and Pearson $r$ is computed per sample then averaged.

\paragraph{CoT generation.}
Reasoning chains are generated with temperature 0.7, \texttt{do\_sample=True}, and a maximum length of 256 tokens. The \texttt{</think>} token marks the end of reasoning. If the model does not produce \texttt{</think>} within 256 tokens, we append it manually.

\paragraph{Masking protocol (Section~\ref{sec:masking}).}
Token-level random masking replaces selected positions with the \texttt{<|endoftext|>} token (token ID 151643, the model's EOS/padding token). Masking is applied at 10 ratios: $\{0\%, 10\%, 20\%, \ldots, 90\%\}$.

\begin{itemize}[leftmargin=*]
    \item \textbf{Mask History}: For each sample, we identify all SID token positions in the user message (before the assistant turn). A random subset of these positions (determined by the mask ratio) is replaced with the padding token. The CoT is generated once from the full history and remains fixed across all mask ratios.
    \item \textbf{Mask CoT}: The full CoT is generated from the unmasked history. A random subset of CoT token positions (excluding the final \texttt{</think>} token) is replaced with the padding token. The history remains intact.
\end{itemize}

Both conditions preserve the original sequence length and positional encodings. A fixed random seed (42) ensures reproducibility, with a separate RNG instance for masking operations.

\paragraph{CoT length experiment (Section~\ref{sec:masking}, Finding 3).}
For each sample, we generate the full reasoning chain (up to 256 tokens) from the unmasked history. We then truncate the CoT at 8 lengths: $\{32, 64, 96, 128, 160, 192, 224, 256\}$ tokens. At each truncation point, the last token is replaced with \texttt{</think>} to maintain a well-formed reasoning boundary. The history remains intact across all truncation lengths. We measure the SID Attention Ratio and drift--bias correlation at each length, using the same attention extraction and logit computation procedures described above.

\paragraph{Statistical significance.}
All reported trends are verified with Wilcoxon signed-rank tests comparing the 0\% and 75\% mask conditions ($p < 0.01$ for all primary metrics on the Product dataset).

\section{LICD Efficiency Analysis}

\label{sec:efficiency}

\paragraph{Motivation.}
LICD adds compression and three-view scoring on top of the \textit{ThinkOn} pipeline.
We quantify the additional inference latency to assess its practical deployability.

\paragraph{Method.}
We estimate latency using the measured generation speed, i.e., 50 tokens/s for 1.7B and 20 tokens/s for 8B, and the measured prefill speed, i.e., 3000 tokens/s for 1.7B and 1500 tokens/s for 8B.
We compare the latency of \textit{ThinkOff}, \textit{ThinkOn}, and our full method.

\begin{figure}[t]
    \centering
    \includegraphics[width=0.85\linewidth]{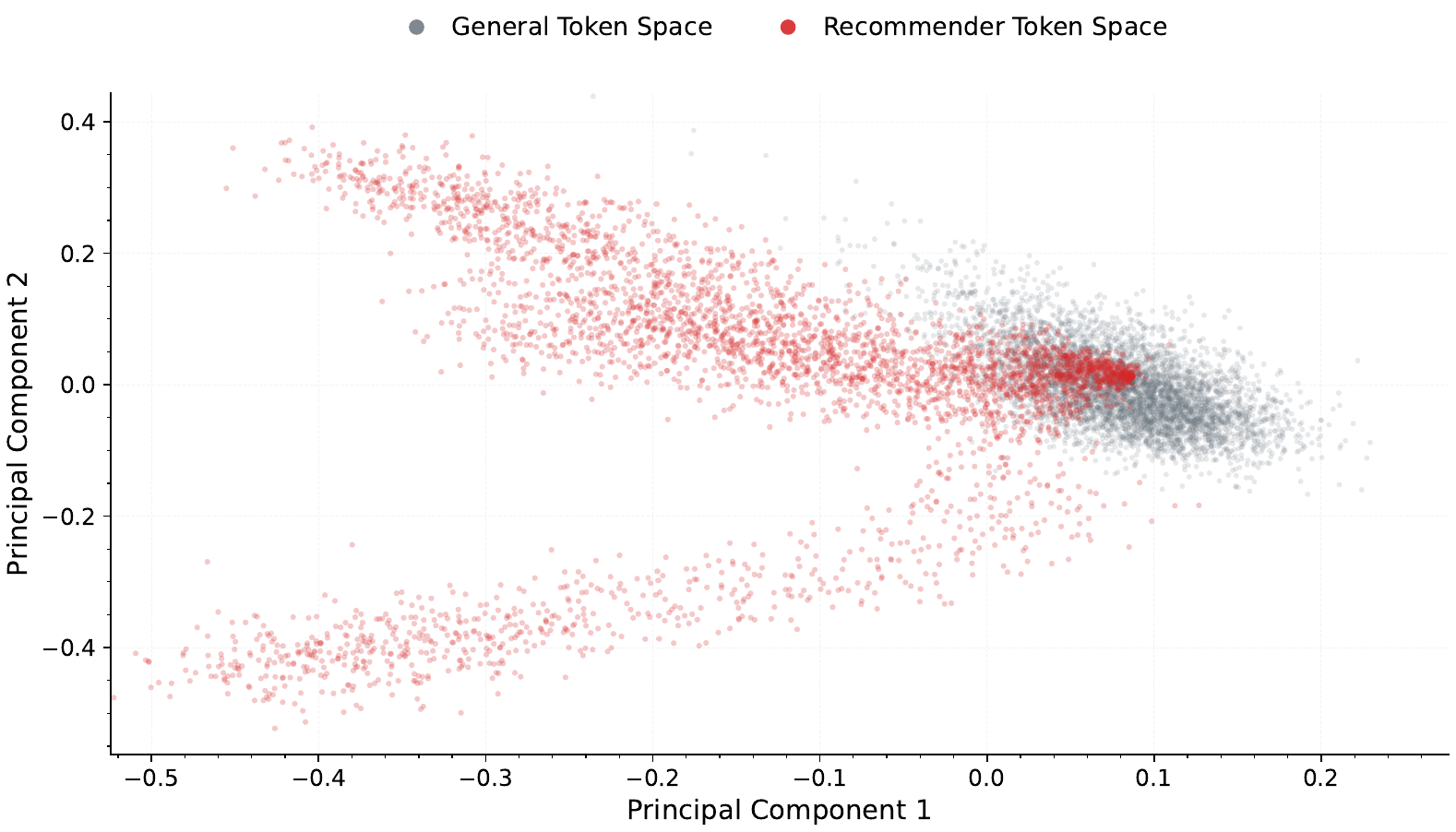}
    \caption{PCA visualization of token embeddings from the Semantic ID Subspace and the General Subspace. The two distributions are partially aligned but remain structurally distinct.}
    \label{fig:pca}
\end{figure}

\begin{table}[t]
\centering
\small
\begin{tabular}{lcccc}
\toprule
& \multicolumn{2}{c}{\textbf{Product}} & \multicolumn{2}{c}{\textbf{AD}} \\
\cmidrule(lr){2-3} \cmidrule(lr){4-5}
\textbf{Method} & \textbf{1.7B} & \textbf{8B} & \textbf{1.7B} & \textbf{8B} \\
\midrule
ThinkOff latency (s) & 3.1  & 7.2  & 2.7  & 6.2  \\
ThinkOn latency (s)  & 26.1 & 64.6 & 26.1 & 64.8 \\
LICD latency (s)     & 31.0 & 74.9 & 30.9 & 74.4 \\
\textbf{Ours / ThinkOn} 
& \textbf{1.19$\times$} 
& \textbf{1.16$\times$} 
& \textbf{1.18$\times$} 
& \textbf{1.15$\times$} \\
\bottomrule
\end{tabular}
\caption{Latency comparison. Ours/ThinkOn denotes the relative latency compared with ThinkOn.}
\label{tab:efficiency}
\end{table}

\paragraph{Analysis.}
% Table~\ref{tab:efficiency} shows that our method introduces only moderate latency overhead over ThinkOn, ranging from 15\% to 19\% across datasets and model scales.
% This is because the dominant cost, CoT generation, is already incurred by the ThinkOn pipeline.
% Our additional cost mainly comes from the compression step and the extra scoring passes used for contrastive reranking.

Table~\ref{tab:efficiency} shows that our method introduces only moderate latency overhead over \textit{ThinkOn}, ranging from 15\% to 19\% across datasets and model scales.
This is because the dominant cost, CoT generation, is already incurred by the \textit{ThinkOn} pipeline.
Our method does not generate longer reasoning chains; instead, it compresses the existing reasoning output and performs additional scoring passes for contrastive reranking.
Therefore, the extra cost mainly comes from the compression step and the prefill computation for the Expert and Amateur views.

\paragraph{Conclusion.}
% Our method achieves better recommendation performance than both ThinkOn and ThinkOff while adding less than 19\% latency over ThinkOn.
% Thus, the proposed correction framework remains practical for inference-time deployment.

Our method achieves better recommendation performance than both \textit{ThinkOn} and \textit{ThinkOff} while adding less than 19\% latency over \textit{ThinkOn}.
This suggests that correcting linguistic inertia does not require a substantially heavier reasoning pipeline; a lightweight compression-and-calibration procedure is sufficient to make reasoning more useful for SID-grounded recommendation.

\section{Subspace Visualization}
\label{app:pca}

To provide intuition for the partial alignment between the Semantic ID Subspace and the General Subspace, we perform Principal Component Analysis (PCA)~\cite{pca1,pca2} on the token embeddings from the model's embedding layer. As shown in Fig.~\ref{fig:pca}, the two subspace distributions exhibit partial overlap---indicating that pre-training achieves some degree of alignment---yet remain semantically distinct. This structural separation explains why relying primarily on general-subspace tokens (e.g., extended CoT) may reduce access to semantic-ID-specific evidence during decoding.

% \section{Attention Heatmap Visualizations}
% \label{app:heatmaps}

% We provide full attention heatmaps for a representative sample under both Think=Off and Think=On settings. These visualizations complement the quantitative metrics presented in Section~\ref{sec:linguistic_inertia} by showing the spatial distribution of attention across the entire input sequence.

% \foreach \p in {1,2,3,4,5,6,7} {
%     \begin{figure*}[t]
%         \centering
%         \includegraphics[width=\textwidth,height=0.78\textheight,keepaspectratio,page=\p]{Sections/Figures/heatmap_off_mean.pdf}
%         \caption{Mean attention heatmap with thinking mode disabled (part \p).}
%         \label{fig:heatmap_off_mean_\p}
%     \end{figure*}
% }

% \foreach \p in {1,2,3,4,5,6,7,8} {
%     \begin{figure*}[t]
%         \centering
%         \includegraphics[width=\textwidth,height=0.78\textheight,keepaspectratio,page=\p]{Sections/Figures/heatmap_on_mean.pdf}
%         \caption{Mean attention heatmap with thinking mode enabled (part \p).}
%         \label{fig:heatmap_on_mean_\p}
%     \end{figure*}
% }

\end{document}